\documentclass[]{aa}  
\usepackage{amsmath}
\usepackage{graphicx}
\usepackage{txfonts}
\usepackage{natbib} 
\bibpunct{(}{)}{;}{a}{}{,} % to follow the A&A style

\usepackage{verbatim}
\usepackage{url}
\usepackage{epstopdf}

\begin{document}

   \title{Impact of measurement errors on the inferred stellar asteroseismic ages.}
   \subtitle{Statistical models for intermediate age main sequence and red giant branch stars}

   \author{G. Valle \inst{1,2,3}, M. Dell'Omodarme \inst{3}, P.G. Prada Moroni
     \inst{2,3}, S. Degl'Innocenti \inst{2,3} 
          }
   \titlerunning{Impact of observational errors on stellar asteroseismic ages}
   \authorrunning{Valle, G. et al.}

   \institute{
INAF - Osservatorio Astronomico di Collurania, Via Maggini, I-64100, Teramo, Italy 
\and
 INFN,
 Sezione di Pisa, Largo Pontecorvo 3, I-56127, Pisa, Italy
\and
Dipartimento di Fisica "Enrico Fermi'',
Universit\`a di Pisa, Largo Pontecorvo 3, I-56127, Pisa, Italy
 }

   \offprints{G. Valle, valle@df.unipi.it}

   \date{Received 27/07/2018; accepted 05/10/2018}

  \abstract
{}
{We aim to perform a theoretical investigation on the direct impact of measurement errors in the observational constraints on the recovered age for stars in main sequence (MS) and red giant branch (RGB) phases. We assumed that a mix of classical (effective temperature $T_{\rm eff}$ and metallicity [Fe/H]) and asteroseismic ($\Delta \nu$ and $\nu_{\rm max}$) constraints were available for the objects.
   }
{       
Artificial stars were sampled from a reference isochrone and subjected to random Gaussian perturbation in their observational constraints to simulate observational errors. The ages of these synthetic objects were then recovered by means of a Monte Carlo Markov chains approach over a grid of pre-computed stellar models. To account for observational uncertainties the grid covers different values of initial helium abundance and mixing-length parameter, that act as nuisance parameters in the age estimation.
}
  {
The obtained differences between the recovered and true ages were modelled against the errors in the observables. This procedure was performed  by means of linear models and projection pursuit regression models. The first class of statistical models provides an easily generalizable result, whose robustness is checked with the second method. 
From linear models we find that no age error source dominates in all the evolutionary phases. Assuming typical observational uncertainties, for MS the most important error source in the reconstructed age is the effective temperature of the star. An offset of 75 K accounts for an underestimation of the stellar age from 0.4 to 0.6 Gyr for initial and terminal MS. An error of 2.5\% in $\nu_{\rm max}$ resulted the second most important source of uncertainty accounting for about $-0.3$ Gyr. The 0.1 dex error in [Fe/H] resulted particularly important only at the end of the MS, producing an age error of $-0.4$ Gyr. For the RGB phase the dominant source of uncertainty is $\nu_{\rm max}$,  causing an underestimation of about 0.6 Gyr; the offset in the effective temperature and $\Delta \nu$ caused respectively an underestimation and overestimation of 0.3 Gyr. 
We find that the inference from the linear model is a good proxy for that from projection pursuit regression models. Therefore, inference from linear models can be safely used thanks to its broader generalizability. 
Finally, we explored the impact on age estimates of adding the luminosity to the previously discussed observational constraints. To this purpose, we assumed -- for computational reasons -- a 2.5\% error in luminosity, much lower than the average error in the Gaia DR2 catalogue. However, even in this optimistic case, the addition of the luminosity does not increase precision of age estimates. Moreover, the luminosity resulted as a major contributor to the variability in the estimated ages, accounting for an error of about $-0.3$ Gyr in the explored evolutionary phases. }
{}


   \maketitle

\section{Introduction}\label{sec:intro}

The problem of obtaining insight into the fundamental, but not directly observable, stellar parameters received growing attention in the last decade, owing to the improvement either in observational data availability and quality and to the development of fast and reliable inversion techniques. These methods adopt large sets of stellar models and estimate the best fundamental parameters by a maximum likelihood or Bayesian approach.

Despite of the popularity of these techniques, only few theoretical works have explored the basis of these approaches, focusing on the evaluation of the biases and random variability expected in the recovery procedure. As an example, \citet{Basu2010, Gai2011, scepter1, eta} shed some light on the performances of inversion techniques based on local interpolation for single stars, while \citet{binary} report an analysis concerning binary systems. The theoretical bias in the recovered overshooting parameter is explored in \citet{overshooting, TZFor}.
Recently \citet{Bellinger2016} has adopted a  machine
learning approach based on random forest to connect observable quantities of stars to other quantities inferred from models. 
\citet{Schneider2017} explores the influence on the maximum likelihood estimates of the correlation among the observables. They conclude that neglecting correlations causes a bias on the  inferred stellar parameters and can alter their precision.
\citet{Angelou2017} study the correlation among classical and asteroseismic observables in order to determine the capacity of each observable to probe structural components of stars and infer their evolutionary histories.

Ultimately, the theoretical understanding of several key points of these models is firmly established. However there are many questions that are still unexplored. A notable example is the quantification of how the errors in the observational constraints -- either from random fluctuations or systematic offsets -- impact on the reconstructed stellar age.
Several studies mentioned above address some aspects related to this question, often 
providing a posterior age distribution obtained by propagating the assumed observational uncertainties through the fitting procedure. However,   
a comprehensive understanding of how much an error in an given observational quantity directly influences the reconstructed age, and how this is linked to the evolutionary stage of the star is still lacking.
The question is indeed of non-negligible relevance, because this knowledge can help to focus the attention on the uncertainty source that most impacts the ages estimates. Refining the accuracy and precision of these error-influence observational constraints will be highly beneficial for age estimation.

This paper aims to partially fill this gap, quantifying the impact of observational errors on the estimated age for stars in the main sequence and in the red giant branch phase. 
We have focussed on the scenario in which classic and asteroseismic constraints are available. In detail, we adopted the effective temperature, the metallicity [Fe/H], the average large frequency spacing $\Delta \nu$ and the frequency of maximum oscillation power $\nu_{\rm max}$, which are measured from most of target stars.
We also explored the case in which the stellar luminosity, provided by Gaia DR2 catalogue \citep{gaiadr2-2018a}, is available as supplementary constraint.

To this purpose one should carefully consider the several input entering in stellar model computations not constrained by the observations. 
These uncertainty sources can be roughly divided into two classes: single and multiple parameters. A notable example of the first class is the initial helium content; since helium lines are detectable in very hot stars, a direct measurement of the helium abundance is in most cases  not achievable. Other examples in the first class stem from the lack of firm theoretical modelling of some precesses, such as the convective core overshooting and the superadiabatic convection. Both these processes are generally included in the stellar computations by free parameters, whose values should be calibrated. All the uncertainties on these parameters can be accounted for by adopting in the fit a grid of stellar models covering a wide range of the possible parameter values. The final age estimate is then marginalized over these nuisance parameters.

The second class poses a bigger problem and no sensible procedure can account for the uncertainty associated with them. Among multi-parameter uncertainty sources, the most important are probably the mean Rosseland opacities, the equation of state, the cross section of the nuclear reactions (for the physics of the models), and the adopted chemical mixture. These quantities are provided to the stellar evolutionary codes by means of pre-computed tables, typically without any associated uncertainty from the literature. A comparison of the tables computed by different authors can shed some light about the variability present in these values, but it also reveals a disparate range of differences for the various zones of the tables \citep{rose2001,OP2005}. Therefore a single parametric scaling -- for example by increasing the values in a table by say 5\% -- could be helpful in providing a rough idea of the relevance of a parameter \citep[see e.g.][]{incertezze1}. It is, however, of no real interest whenever one tries to obtain a reliable estimate of the error propagation on final fitted parameter.  
As a consequence, the uncertainty owing to these multi-parameter sources is usually neglected and the error propagation on the final estimate is usually provided by fixing the input physic. 
Given these difficulties, in this work we have adopted a pragmatic approach focusing only on the first class of uncertainty sources.  
    
The study is organized in the following way. In Sect.~\ref{sec:method} we describe the procedure adopted for the investigation, the grid of stellar models, and the approach to the age estimation through observations inversion. The statistical models to analyse the impact of the observational errors on the final age estimate are presented and discussed in Sect. \ref{sec:results}. Results of adding the luminosity constraints provided by Gaia are discussed in Sect. \ref{sec:gaia}.
We give some concluding remarks in Sect. \ref{sec:conclusions}.

\section{Methods}\label{sec:method}

All the investigations were performed by assuming a reference scenario, with a fixed chemical composition and age. In the reference scenario we also fixed the efficiency of superadiabatic convection by adopting a value of the mixing-length parameter  $\alpha_{\rm ml} = 1.74$, which corresponds to the solar calibrated value.
The reference case chemical composition approximatively mimics the old cluster NGC6791 having a metallicity $Z = 0.0267$ and an initial helium value $Y = 0.275$, corresponding to a helium-to-metal enrichment ratio $\Delta Y/\Delta Z = 2.0$. This choice implies [Fe/H] = 0.3, assuming the solar heavy-element mixture by \citet{AGSS09}. The reference age was chosen to be 7.5 Gyr.

A dense grid of stellar models was computed around the reference scenario. In fact, to obtain a sensible error estimate of the fitted age the stellar grid should span an appropriate range not only in mass (or age for isochrones) but also in other quantities that impact the stellar evolution, such as the initial chemical composition ($Z$ and $Y$), and in parameters that affect the position of the red giant branch (RGB), such as the value of the mixing-length parameter $\alpha_{\rm ml}$. The grid of stellar models is presented in Sect.~\ref{sec:grids}.

\begin{figure*}
        \centering
        \includegraphics[height=17cm,angle=-90]{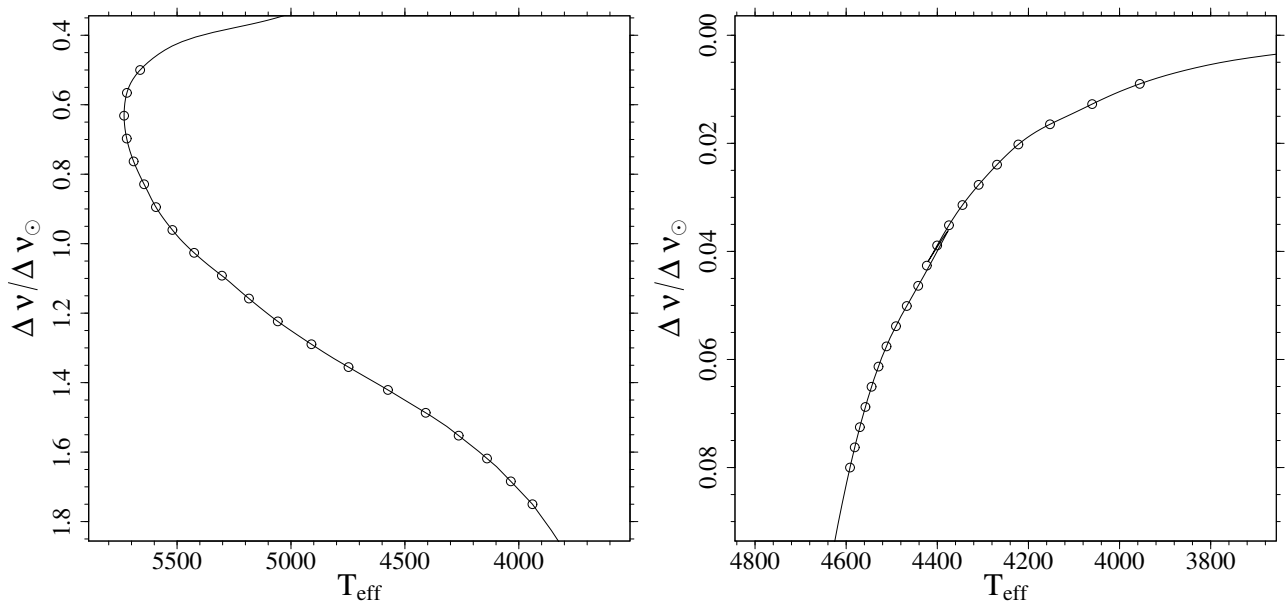}
        \caption{{\it Left}: Main sequence of the reference isochrone at 7.5 Gyr ($Z = 0.02674$, $Y = 0.2752$, $\alpha_{\rm ml} = 1.74$) in the $T_{\rm eff}$ -- $\Delta \nu$ plane. The circles mark the position of the synthetic stars adopted for the analysis. {\it Right}: As in left panel but for the RGB phase.
        }
        \label{fig:iso-ref}
\end{figure*}

From the reference isochrone we sampled $n = 20$ stars from both the main sequence (MS)
 and the RGB stages.
The stars were chosen by sampling uniformly in $\Delta \nu$ the MS ($\Delta \nu \in [1.75, 0.5]$ $\Delta \nu_{\sun}$) and the RGB ($\Delta \nu \in [0.08, 0.009]$ $\Delta \nu_{\sun}$)).
The choice implies a mass range [0.6, 1.1] $M_{\sun}$ in MS and [1.15, 1.16] $M_{\sun}$ in RGB.
Figure~\ref{fig:iso-ref} shows the reference isochrone with the positions of the synthetic stars considered in the analysis  superimposed. 

The observables of each synthetic star were then subjected to a Gaussian perturbations, mimicking the observational errors. The assumed uncertainties are: 75 K in $T_{\rm eff}$, 0.1 dex in [Fe/H], 1\% in  $\Delta \nu$, and 2.5\% in $\nu_{\rm max}$. The errors in the asteroseismic quantities were chosen 
taking into account the values quoted in SAGA of 0.7\% and 1.7\%  on $\Delta \nu$ and 
$\nu_{\rm max}$ \citep{Casagrande2014} and those in the APOKASC catalogue
of 2.2\% and 2.7\% \citep{Pinsonneault2014}. However the exact value of the uncertainties is not of paramount importance, because the aim is not to obtain a posterior distribution for the reconstructed ages but to study the age dependence on the observables. Therefore it is only relevant that the errors cover a sensible range around the reference value. 

Finally, ages for all the perturbed stars were recovered  by means of pure geometrical isochrone fitting by Monte Carlo Markov chain (MCMC) simulations (Sect.~\ref{sec:fittingML}). The median value of the recovered age was stored along with the perturbation of the four observables adopted in the fit. The whole procedure was repeated $N = 200$ times to assess the importance of the measurement errors. The adopted value of $N$ was directly verified to be large enough to reduce the variability on the result to a negligible value. A final dataset of $n \times N = 4000$ synthetic data was then subjected to statistical modelling by regressing the error in the fitted age against the errors in the observational variables. Statistical models were then adopted to evaluate the effects of the observational constraints on the recovered stellar ages.

\subsection{Stellar model grid}
\label{sec:grids}

The model grid was computed for masses in the range [0.4, 1.3] $M_{\sun}$, with a step of 0.05 $M_{\sun}$.
The initial metallicity [Fe/H] was varied from 0.15 dex to 0.45 dex, with
a step of 0.05 dex. 
The solar heavy-element mixture by \citet{AGSS09} was adopted. 
Nine initial helium abundances were considered at fixed metallicity by adopting the commonly used
linear relation $Y = Y_p+\frac{\Delta Y}{\Delta Z} Z$
with the primordial abundance $Y_p = 0.2485$ from WMAP
\citep{peimbert07a,peimbert07b} and with a helium-to-metal enrichment ratio $\Delta Y/\Delta Z$
in the range [1, 3] with a grid step of 0.25 \citep{gennaro10}. 

The FRANEC code \citep{scilla2008, Tognelli2011} was used to compute the stellar models, in the same
configuration as was adopted to compute the Pisa Stellar
Evolution Data Base\footnote{\url{http://astro.df.unipi.it/stellar-models/}} 
for low-mass stars \citep{database2012}. 
The models were computed
for five different values of mixing-length parameter $\alpha_{\rm ml}$ in the range [1.54, 1.94], with a step of 0.1, 1.74 being the solar-scaled value\footnote{The calibration is performed repeating the Sun evolution by changing $Z$, $Y$ and $\alpha_{\rm ml}$. The iteration stops when, at the Sun age, the computed radius, luminosity, effective temperature, and photospheric [Fe/H] match the observed values with relative tolerance $10^{-4}$.}. Microscopic diffusion is considered according to \citet{thoul94}.
Further details on the stellar models can be found in \citet{cefeidi,eta,binary} and references therein. Isochrones, in the age range [5, 10] Gyr, with time steps of 100 Myr,  were computed according to the procedure described in 
\citet{database2012, stellar}.

\subsection{Fitting procedures}\label{sec:fittingML}

For age recovery the widely adopted pure geometrical isochrone fitting method was employed \citep[see][and references therein]{Valls2014}, as described in \citet{geoML}. 
For readers' convenience we summarize here the method.
Let $\theta = (\alpha_{\rm ml}, \Delta Y/\Delta Z, Z, {\rm age})$ be the vector of parameters characterizing the isochrones and $q \equiv \{T_{\rm eff}, {\rm [Fe/H]}, \Delta \nu, \nu_{\rm max}\}_i$ the vector of observed quantities for a star. Let $\sigma$ be the vector of observational uncertainties.
For each point $j$ on a given isochrone we defined $q_j(\theta)$ the vector of theoretical values. Finally we computed the geometrical distance $d_{j}(\theta)$ between the observed star and the $j$th point on the isochrone, defined as
\begin{equation}
d_{j}(\theta) = \left\lVert \frac{q - q_j(\theta)}{\sigma} \right\rVert. \label{eq:dist}
\end{equation} 
Then the statistic:
\begin{equation}
\chi^2(\theta) = \min_j d_{j}^2(\theta)
\end{equation} 
can be adopted to compute the probability $P(\theta)$ that the synthetic star comes from the given isochrone. 
In fact, assuming an independent Gaussian error model, we have
\begin{equation}
P(\theta) \propto \exp(-\chi^2/2) \label{eq:prob-geom},
\end{equation} 
where we neglected the normalization constant.
The method suffers for an intrinsic degeneracy -- such as the age-metallicity degeneracy -- and different sets of parameters  can provide similar likelihood \citep[see][and references therein]{Valls2014}. 
A possible proposed remedy is to factorize in the likelihood computation the evolutionary time step (see e.g. the discussion in \citealt{Valls2014, geoML}), but direct verification for RGB phase showed that this alternative approach has its own difficulties because it can provide biased estimates \citep{geoML}.

The age estimation  was performed adopting an MCMC process, using the likelihood in Eq.~(\ref{eq:prob-geom}). The method has been widely adopted in the literature since the original formulation by \citet{Metropolis1953} and \citet{Hastings1970}. Briefly, one starts by establishing a good initial point in the $\theta$ space and by evaluating the scale over which the likelihood function varies. Such estimates were obtained by random sampling 2\,000 values in $\theta$ space spanned by the isochrone grid, and evaluating the relative likelihood. 
This step sets the initial value and the initial guess of the covariance matrix in the $\theta$ space for the jump function. Then a burning-in chain of 3\,000 points is generated in the following way.
Let us call $\theta_k$ the value of the parameters after the step $k $ and $\Sigma$ the covariance matrix. A proposal point in the $\theta$ space is generated adopting a Gaussian jump function with mean $\theta_k$ and covariance $\Sigma$
\begin{equation}
\theta_{k+1} \sim N(\theta_k, \Sigma)\label{eq:MCMC-jump}
\end{equation} 
Let us call $P(\theta)_k$ and $P(\theta)_{k+1}$ the likelihood of the solutions $\theta_k$  and 
$\theta_{k+1}$. We define $P_r = \min(\frac{P(\theta)_{k+1}}{P(\theta)_{k}},1)$. Then the following chain rule applies:
\begin{equation}
\theta_{k+1} = \begin{cases} \theta_{k+1}, & \mbox{with probability } P_r \\ \theta_{k}, & \mbox{with probability } 1-P_r \end{cases}
\end{equation}
Due to the intrinsic degeneracy among the parameters $\theta$ the convergence speed of the chain is known to be sub-optimal (see \citealt{Haario2001}, and also \citealt{Kirkby-Kent2016} for a specific discussion in a detached binary system fit). Therefore after the burn-in phase the covariance matrix of the last 50\% of the proposed solutions is adopted to transform the $\theta$ variables to $\theta_\perp$ orthogonal ones. The step is performed by means of a principal component analysis \citep{Feigelson2012,simar}, a statistical technique that computes a set of independent and orthogonal linear combinations $\theta_\perp$ of the original $\theta$ variables. The chain is then built in the newly computed space, achieving a better convergence speed. Finally, the transformation is inverted and the results are remapped in the original $\theta$ space.

After the burning-in stage, the MCMC sampling consisted of eight chains of length 5\,000. The length of the chain is sufficient to achieve good convergence and mixing, according to the statistical tests of     
Gelman-Rubin and Geweke \citep{Gelman1992, Geweke1992}. 

\section{Results}
\label{sec:results}

The two datasets (one for MS and one for RGB stars) resulting from the fit of the 4\,000 synthetic stars in both evolutionary stages  were subjected to statistical modelling. 
The analysis was focussed on finding two multi-variate linear models with the age error (estimated minus true age) as dependent variable and the perturbations on the observables as predictors (perturbed values minus real ones). 
The models were then adopted to identify the contribution of the variation of each predictors to the recovered age. In this investigation we have to balance two opposite goals. First, we needed to identify  a model that fits the data with some margin for improvements but adopting simple relations -- such as only first degree predictors. Second,  the inference from this model should be robust.
While the latter aim come from obvious considerations, 
the former one was required because a model with only first-degree predictors and without interactions among them can be used for evaluating the impact of different observational errors $\sigma_1$, by a simple scaling of the contributions with the ratio of $\sigma_1/\sigma$. In particular, the analysis was focussed on searching for models with  no interactions among predictors; this condition  guarantees that the impact of each predictor can be evaluated independently from the values of the others.  

In the following we adopt the error function $\cal E(.)$ to indicate the difference between the true and perturbed value. Therefore ${\cal E}({\rm age})$ is the difference between the recovered and the true age, while ${\cal E}(T_{\rm eff})$ is the perturbation of the effective temperature.
The errors in the asteroseismic quantities enter in the model as a percentage, as in the Monte Carlo simulations.

\subsection{Linear models}\label{sec:linmod}

\begin{figure}
	\centering
	\includegraphics[height=8.2cm,angle=-90]{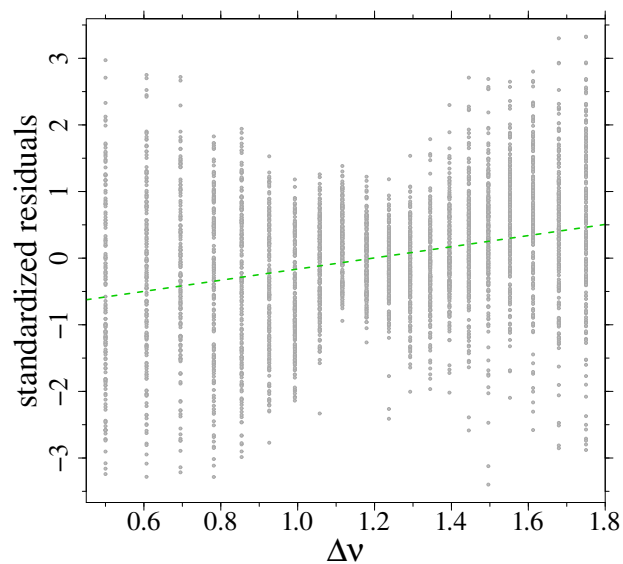}
	\caption{Residual of the linear model in the MS as a function of $\Delta \nu$. The dashed line shows a linear regression of the standardized residuals versus $\Delta \nu$.}
	\label{fig:resid-trend}
\end{figure}

In the search for the simplest models able to describe the data, the analyses for MS and RGB phases were begun by assessing the performance of linear models containing only the four predictors and no interactions of high power terms. However, the analysis of these models highlighted the presence of a trend in the model residuals versus the MS position and a butterfly-shaped behaviour. Figure~\ref{fig:resid-trend} shows the MS model standardized residuals as a function of the $\Delta \nu$ value. The line in the figure is a linear regression between these variables and evidences a significant trend. This trend, and the large inhomogeneity of the estimated residual variance (much larger at the $\Delta \nu$ edges with respect to the centre) are characteristic signs of the presence of interactions among variables unaccounted by the model. As a result, the simple model is not equally effective for the whole MS phase, and a corrective term should be inserted.

\begin{table*}[ht]
	\centering
	\caption{Fit of the least square MS model.} 
	\label{tab:MS-add}
	\begin{tabular}{lrrrr}
		\hline\hline
		& Estimate & Std. error & t value & P value \\ 
		\hline
		Intercept & $-1.43 \times 10^{-1}$ & $6.03 \times 10^{-3}$ & -23.70 & $< 10^{-16}$ \\ 
		$\Delta \nu$ & $1.32 \times 10^{-1}$ & $4.84 \times 10^{-3}$ & 27.35 & $< 10^{-16}$ \\ 
		${\cal E}(T_{\rm eff})$ & $-9.48 \times 10^{-3}$ & $8.30 \times 10^{-5}$ & -114.17 & $< 10^{-16}$ \\ 
		${\cal E}({\rm [Fe/H]})$ & $-4.92 \times 10^{0}$ & $6.29 \times 10^{-2}$ & -78.14 & $< 10^{-16}$ \\ 
		${\cal E}(\Delta \nu)$ & $2.48 \times 10^{1}$ & $6.20 \times 10^{-1}$ & 39.97 & $< 10^{-16}$ \\ 
		${\cal E}(\nu_{\rm max})$ & $-1.43 \times 10^{1}$ & $2.53 \times 10^{-1}$ & -56.59 & $< 10^{-16}$ \\ 
		$\Delta \nu \times {\cal E}(T_{\rm eff})$ & $2.25 \times 10^{-3}$ & $6.62 \times 10^{-5}$ & 33.97 & $< 10^{-16}$ \\ 
		$\Delta \nu \times {\cal E}({\rm [Fe/H]})$ & $2.01 \times 10^{0}$ & $5.01 \times 10^{-2}$ & 40.07 & $< 10^{-16}$ \\ 
		$\Delta \nu \times {\cal E}(\Delta \nu)$ & $-1.90 \times 10^{1}$ & $4.98 \times 10^{-1}$ & -38.13 & $< 10^{-16}$ \\ 
		$\Delta \nu \times {\cal E}(\nu_{\rm max})$ & $2.64 \times 10^{0}$ & $2.02 \times 10^{-1}$ & 13.05 & $< 10^{-16}$ \\ 
		\hline
	\end{tabular}
	\tablefoot{The columns report the estimate of the coefficients, their statistical standard error, the $t$ value (obtaining dividing the first column by the second), and finally the $P$ value for the significance. A $P$ value lower than 0.05 is generally adopted as statistical significance threshold.}
\end{table*}

Thus the MS model was adjusted by adding an explicit dependence on the $\Delta \nu$ value, which decreases during the MS evolution, providing a proxy for the evolutionary stage. Moreover, the interactions among predictors and  $\Delta \nu$ were required. The final linear model for the MS phase is:
\begin{equation}
{\cal E}({\rm age}) \, ({\rm Gyr}) = \Delta \nu * \left({\cal E}(T_{\rm eff}) + {\cal E}({\rm [Fe/H]}) + {\cal E}(\Delta \nu) + {\cal E}(\nu_{\rm max}) \right),
\label{eq:MS-model}
\end{equation}
where we adopt the operator "*" to mark the interaction between predictors: an interaction term $A * B$ is interpreted as $A + B + A \times B$.
The explicit dependence on the fit coefficients was not expressed in the model to keep the notation compact and can be found in Appendix~\ref{app:fullform}. 

The coefficients resulting from the least squares fit are reported in Table~\ref{tab:MS-add}.
Adopting these coefficients, it is possible to extract the quantity of direct interest in the present investigation, in other words, the impact on the estimated age of a $1 \sigma$ offset in the predictors. We note that the models apply either to observational random errors and to systematic errors. This is particularly relevant since it is well known that 
there is an approximately 5\% offset between the scaling relation masses and the true masses of giants \citep[see e.g.][]{Guggenberger2017,Themessl2018} which reflects in incorrect ages. Moreover
one of the predictors, namely the effective temperature, is subject to a non-negligible systematic uncertainty \citep[see e.g.][]{Ramirez2005,Schmidt2016}. By multiplying the coefficients of the fit by the synthetic observational error vector $\sigma$ it is possible to evaluate the impact of $1 \sigma$ observational error. Due to the dependence of the model on $\Delta \nu$, the computations were performed twice: first for a value $\Delta \nu$ = 1.75 $\Delta \nu_{\sun}$ (initial MS phase, see Fig.~\ref{fig:iso-ref}), and second for $\Delta \nu$ = 0.50 $\Delta \nu_{\sun}$ (end of MS phase). The impacts on the age (in Gyr) are reported in the first two rows of Table~\ref{tab:contributi}. 

For the MS, the main contribution to the age error is the uncertainty on the effective temperature: an increase of $T_{\rm eff}$ by 75 K offsets the age estimate by $-0.42$ Gyr in the initial part of the MS, and of $-0.62$ Gyr in its terminal part, that is, from about 6\% to 8\% of the nominal age of 7.5 Gyr. The second most important contributor is the error in $\nu_{\rm max}$ (2.5\%), accounting for an offset from $-0.24$ Gyr to $-0.33$ Gyr.
The $1 \sigma$ error in the metallicity (0.1 dex) is particularly relevant at the MS end, when it accounts for $-0.39$ Gyr (about two thirds of the impact of the effective temperature error), while it is only $-0.14$ Gyr in the first part of the MS. Finally the effect on the age determination of the error in $\Delta \nu$ gave the lowest impact, from $-0.08$ Gyr to $+0.15$ Gyr.

\begin{figure*}
        \centering
        \includegraphics[height=17cm,angle=-90]{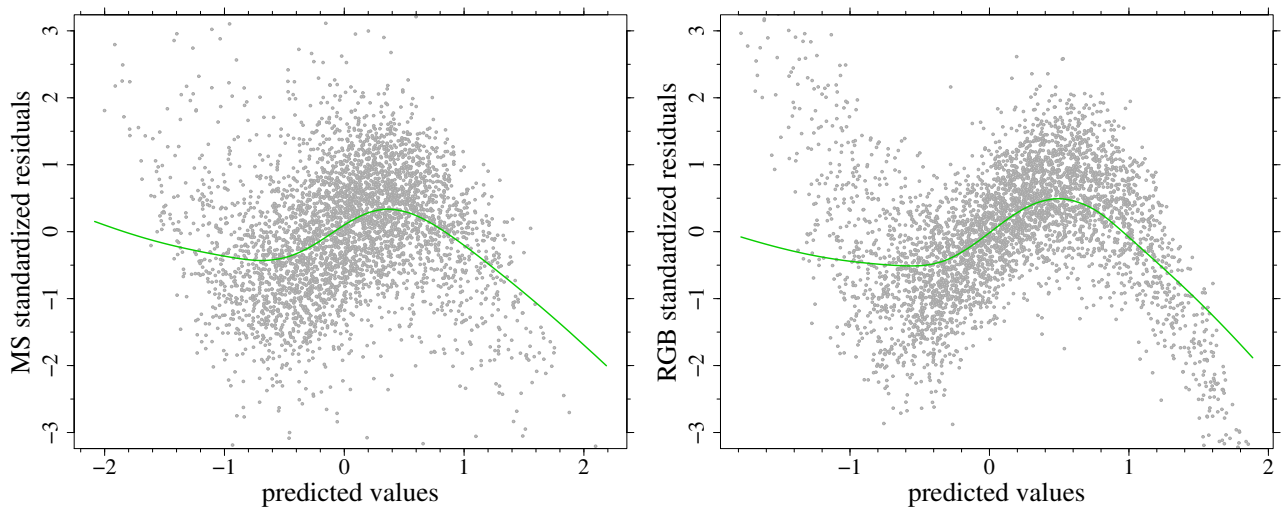}
        \caption{{\it Left}: Residuals of the linear model for the age of MS stars as a function of the predicted values. The solid line shows a smoother of the data obtained by the LOESS technique. {\it Right}: As in the {\it left} panel, but for RGB stars.}
        \label{fig:resid}
\end{figure*}

The assessment of the model validity, however, reveals some problems.
The inspection of the residual plot versus the fitted values (left panel in Fig.~\ref{fig:resid}) -- a standard check for linear models -- shows that some information is retained in the data. The trend is clearly seen in the solid line -- a smoother of the data obtained by the LOESS (LOcal regrESSion) technique \citep{Cleveland1988, Cleveland1991}. The clear arc-shaped paths, with heavy variations at the wings present in both panels is a typical signature of interaction among predictors not accounted for in the model.
Adding higher-order terms and interactions to improve the models did not cleanly solved the issue, while the complexity of the models increased very much. 
Moreover, the aim of the work is to develop models te be used to quantify the impact of different assumptions of the observational errors. In the lack of interactions and high-order terms, this can be simply obtained  by scaling the values reported in Table~\ref{tab:contributi} by the ratio of the new versus adopted $\sigma$ values. On the contrary, in the presence of interactions, the impact of a parameter depends also on the values of the others, making the impact estimation a bit more complex.
 
Ultimately, we found that the exact form of interaction between the predictors is extremely cumbersome and required the adoption of a much more robust and versatile statistical technique to be accounted for. These models are discussed in Sect.~\ref{sec:ppr}, but happily the inference on the impact of the single error source, presented in this section, was found to be a very good proxy for that from the more complex models.
 
 \begin{table*}[ht]
        \centering
        \caption{As in Table~\ref{tab:MS-add}, but for RGB linear model.}
        \label{tab:RGB-add}
        \begin{tabular}{lrrrr}
                \hline\hline
                & Estimate & Std. error & t value & P value \\ 
                \hline
                Intercept & $1.68 \times 10^{-1}$ & $1.37 \times 10^{-3}$ & 122.73 & $< 10^{-16}$ \\ 
                ${\cal E}(T_{\rm eff})$ & $-3.55 \times 10^{-3}$ & $1.85 \times 10^{-5}$ & -191.81 & $< 10^{-16}$ \\ 
                ${\cal E}({\rm [Fe/H]})$ & $2.84 \times 10^{-1}$ & $1.38 \times 10^{-2}$ & 20.54 & $< 10^{-16}$ \\ 
                ${\cal E}(\Delta \nu)$ & $3.11 \times 10^{1}$ & $1.41 \times 10^{-1}$ & 221.44 & $< 10^{-16}$ \\ 
                ${\cal E}(\nu_{\rm max})$ & $-2.30 \times 10^{1}$ & $5.65 \times 10^{-2}$ & -408.18 & $< 10^{-16}$ \\ 
                \hline
        \end{tabular}
 \end{table*}

The analysis of the RGB phase was performed in a similar way. The adopted linear model was
\begin{equation}
{\cal E}({\rm age}) \, ({\rm Gyr}) = {\cal E}(T_{\rm eff}) + {\cal E}({\rm [Fe/H]}) + {\cal E}(\Delta \nu) + {\cal E}(\nu_{\rm max}).
\label{eq:RGB-model}
\end{equation}
The coefficients of the fit are reported in Table~\ref{tab:RGB-add}, while the impact of the $1 \sigma$ errors are in the last row of Table~\ref{tab:contributi}.
The most important uncertainty source in the recovered age is the error in $\nu_{\rm max}$: an increase  of $\nu_{\rm max}$ by 2.5\% causes an age change of $-0.58$ Gyr  (about 8\% of the true age of 7.5 Gyr). An effective temperature increase of 75 K leads to an age offset of $-0.27$ Gyr, while an increase of 1\% of $\Delta \nu$ changes the age estimation by 0.31 Gyr. The contribution of the 0.1 dex metallicity error resulted negligible.

As for the MS model, the residuals plot for RGB stars (right panel in Fig.~\ref{fig:resid}) also shows a clear path, suggesting that higher-order terms are required to properly account for the dependence of the age error on the predictors. Moreover, as for the MS phase, the adoption of high-order interaction terms does not pose a clear remedy to the problem but adding more complexity.

In summary, both MS and RGB linear models provide a very tight fit to the data. In fact the residual error from the MS model is $\sigma_{\rm MS} = 0.1$ Gyr on a reference age of 7.5 Gyr, with a multiple correlation coefficient $R^2 = 0.97$, implying that 97\% of the variability present in the data is correctly accounted by the model.  
For the RGB model, the residual variance is $\sigma_{\rm MS} = 0.08$ Gyr and $R^2 = 0.98$. 
Thus, the models presented so far appear globally reliable.
To explore the problem of the trends in the residuals (Fig.~\ref{fig:resid}), more complex statistical models is adopted in the following section.

\begin{table*}[ht]
        \centering
        \caption{Contribution to the age error (in Gyr) of a $1 \sigma$ shift of the four predictors adopted in the linear models, in different evolutionary stages.} 
        \label{tab:contributi}
        \begin{tabular}{lrrrrr}
                \hline\hline
                & ${\cal E}(T_{\rm eff})$ & ${\cal E}({\rm [Fe/H]})$ & ${\cal E}(\Delta \nu)$ & ${\cal E}(\nu_{\rm max})$ & $\Delta \nu/\Delta \nu_{\sun}$\\ 
                \hline
                Initial MS & -0.42 & -0.14 & -0.08 & -0.24 & 1.75 \\ 
                End MS & -0.63 & -0.39 & 0.15 & -0.33 & 0.50 \\ 
                RGB & -0.27 & 0.03 & 0.31 & -0.58 & - \\ 
                \hline
        \end{tabular}
\end{table*}

\subsection{Projection pursuit regression models}\label{sec:ppr}

The residual plots in Fig.~\ref{fig:resid} showed the presence of some interactions among the predictors, that cannot be neatly accounted for in the framework of linear models, even adopting higher order interactions. A much more versatile, although much more complex, statistical model is able to deal well with this problem. The inference from this method can be compared with that from linear models of the previous section to evaluate the improvement obtained with the more complex method.
 
To this aim, we adapted two projection pursuit regression (PPR) models to the data. The method is computationally intensive but allows us to deal  -- although not explicitly -- with complex interactions among predictors. This statistical model is a generalization of the additive model and projects the data matrix of the predictors (i.e. the errors in the observable constraints) in the optimal direction before applying smoothing functions to these explanatory variables \citep[see e.g.][]{ppr,venables2002modern,simar}.  The dimension $M$ of the projection subspace is chosen by the modeller. In detail, let $n$ be the sample size, $\mathbf X$ the matrix of the predictors (in our case, the offset in the observational constraints), and $y$ the response variable (the age error). The PPR model is then
\begin{equation} 
y_i = \beta_0 + \sum_{j=1}^M \beta_{ij} \, f_j(\alpha_j^T \mathbf X) + \varepsilon_i 
\quad i=1,\ldots,n
\label{eq:ppr1}
\end{equation}
where the vectors $\alpha_j$ are the projecting directions. The $f_j$ functions are empirically determined smooth functions, usually called ridge functions. $\beta_{ij}$ are the coefficients to be estimated and $\varepsilon_i$ the model residuals. The value of $M$ is generally determined through cross-validation, adopting an iterative stepwise method that stops when the model fit no longer improves.

In summary, a PPR model is similar to an additive model but with the additional projection stage so that it fits the model  $\alpha^T \mathbf X$ rather than the actual predictor $\mathbf X$. The directions $\alpha_j$ are chosen to optimize model fit and constrained to unit length. The ridge functions are estimated using some smoothing methods. The drawback of this flexibility and generality is that it is difficult to interpret the fitted model because  each input variable enters into the model in a complex  way. Thus the model is far more useful for prediction than for understanding the relations among data.
Moreover, a PPR model -- like many deep learning methods -- suffers from a difficult that limit the adoption of a proposed model from different researchers \citep{Ripley1996,venables2002modern, simar}. In fact, it
depends on the empirically determined $f_j$ smoothing functions, which have no analytic expressions. Thus the use of the PPR model is limited by the need to know the $f_j$ functional forms exactly.     
The models were fitted adopting the {\it ppr} function in the MASS package of R 3.5.0 \citep{R, venables2002modern}.

The model for the MS phase adopts as predictor the offsets in the observational constraints and the $\Delta \nu$ value, while the latter is not required in the RGB model. For the MS model a total of $M = 3$ functions were retained in the final model, while $M = 2$ was sufficient for the RGB phase. 
Interested reader can found the coefficients of the $\alpha$ vectors and the empirical smoothing functions in Appendix~\ref{app:ppr}.

\begin{figure*}
        \centering
        \includegraphics[height=17cm,angle=-90]{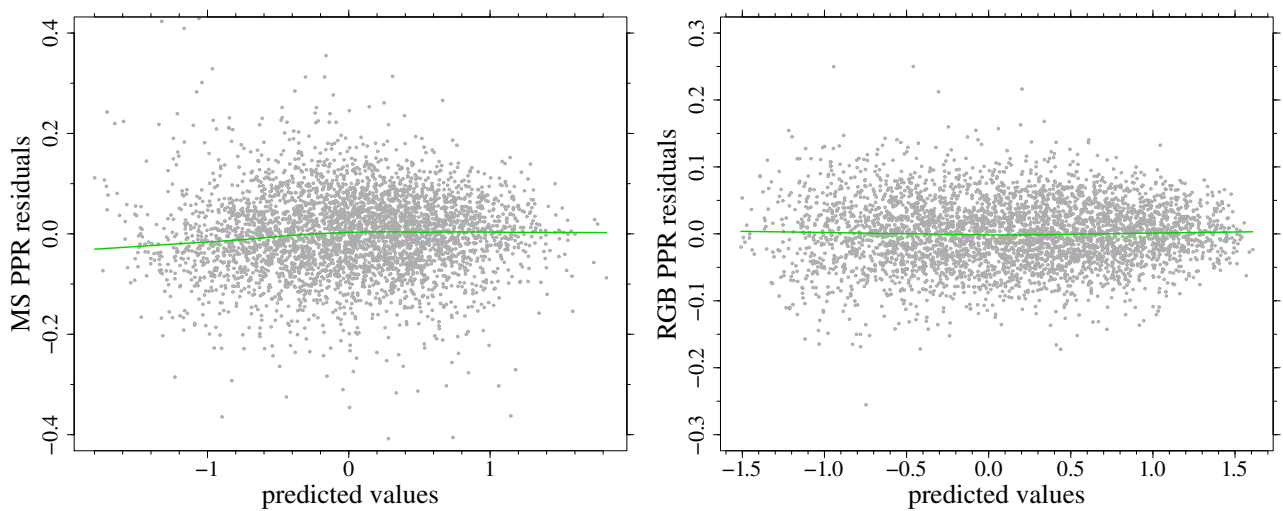}
        \caption{As in Fig.~\ref{fig:resid}, but for the PPR models.}
        \label{fig:resid-ppr}
\end{figure*}

The capability of the PPR model to account for the internal interactions is shown in the residual plots of Fig.~\ref{fig:resid-ppr}. Both panels show that the models succeed in removing the trends resulting from the linear models for MS and RGB phase. The standard deviations of the residuals for the MS and RGB models are about 0.09 Gyr and 0.05 Gyr respectively, a remarkable precision being about 1\% of the true age 7.5 Gyr.

Due to the correct modelling in the whole range of the fitted values, the inference from the PPR model is more robust than that from the linear model. To judge the relevance of this difference we evaluated the impact on the recovered age of the observable uncertainties from PPR models and compared them to those in Table~\ref{tab:contributi}. A non-negligible systematic difference would suggest that the inference from linear model is not adequate to describe the dependence of the age error on the observables offset. As a difference from the simple method of Sect.~\ref{sec:linmod}, the computation of the impact from PPR model is somewhat cumbersome. Due to the presence of hidden interactions, the impact of a predictor shift depends also on the values of the others so we need to evaluate the impact of a single predictor given different values of all the others. To this aim, we constructed a matrix of artificial predictors where the predictors $X_j$ can take only two values: $X_j = \pm \sigma_j$ ($\sigma_j$ is the observational uncertainty of the $j$-th predictor). Then the predictions $\hat y$ of the age shift were obtained from the computed PPR models in Eq.~(\ref{eq:ppr1}) adopting these artificial $\mathbf X$ predictors. Then, for the $i$-th predictor, we computed the difference between the values of $\hat y$ having equal predictor but $i$ and the reference value for all unperturbed predictors. The sign of  perturbations for a $-\sigma$ offset was changed. Finally the range of these differences was computed providing the impact of the $i$-th predictor on the age error. 

\begin{table*}[ht]
        \centering
        \caption{As in Table~\ref{tab:contributi}, but for PPR models.} 
        \label{tab:contributi-ppr}
        \begin{tabular}{lcccc}
        \hline\hline
                & ${\cal E}(T_{\rm eff})$ & ${\cal E}({\rm [Fe/H]})$ & ${\cal E}(\Delta \nu)$ & ${\cal E}(\nu_{\rm max})$\\ 
                \hline
                Initial MS & [-0.48,-0.4] & [-0.15,-0.08] & [-0.11,-0.08] & [-0.27,-0.21] \\ 
                End MS & [-0.68,-0.53] & [-0.43,-0.33] & [0.11,0.16] & [-0.37,-0.29] \\ 
                RGB & [-0.32,-0.26] & [0,0.07] & [0.3,0.37] & [-0.69,-0.6] \\ 
                \hline
        \end{tabular}
\end{table*}

\begin{figure*}
        \centering
        \includegraphics[height=18cm,angle=-90]{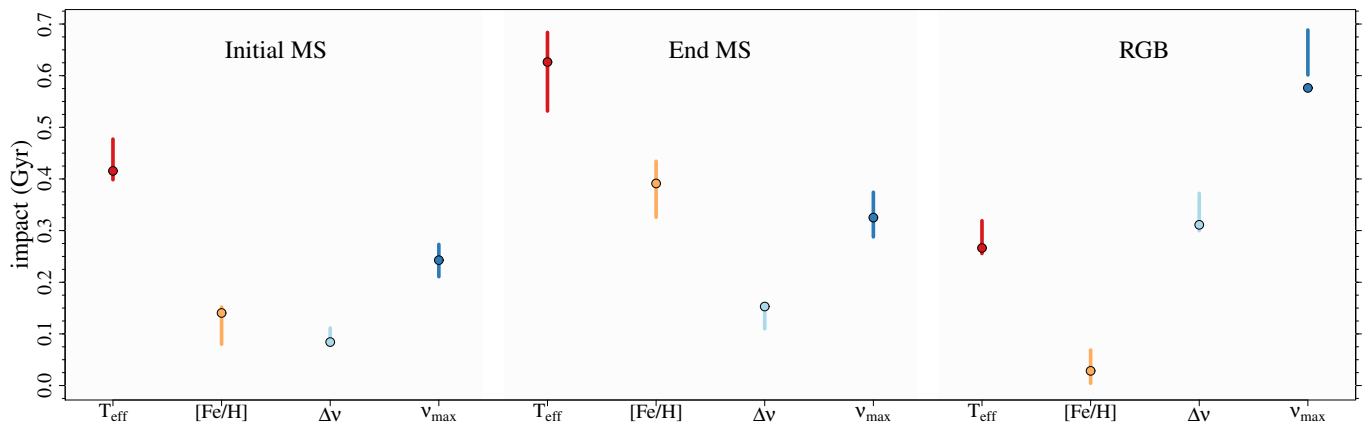}
        \caption{Impact on the estimated age (in Gyr) of the $1 \sigma$ error in the predictors, classified according to the evolutionary stage (from low MS to RGB) and predictors ($T_{\rm eff}$, [Fe/H], $\Delta \nu$, and $\nu_{\rm max}$). The segments identify the impact range estimated from PPR models, while the circles those from linear models (see text).}
        \label{fig:impact}
\end{figure*}

The result of this computation is presented in Table~\ref{tab:contributi-ppr} and in Fig.~\ref{fig:impact}, which also reports the impact estimates for linear models. For a better evaluation of the relevance of the uncertainty, all the impacts in the figure are reported in absolute value. It appears that linear and PPR models estimates are well consistent, because the linear model estimated impact always lie within the range of PPR models, with the only exception of the $\nu_{\rm max}$ error in the RGB phase. Moreover, the impact ranges for PPR models are generally smaller than the effect on the age for the different predictors, allowing to firmly establish what are the major sources of uncertainty.

In summary, it seems that the impact on the age error from linear models are a good proxy for the more precise results from PPR. This is a particularly pleasant finding; due to their simple nature, and to the possibility to directly generalize to different observational uncertainties, the results from the linear model can have a broader scope of applicability. 

\section{Effect of adding the luminosity to the observational constraints}\label{sec:gaia}

The recent advent of the Gaia satellite \citep{gaia2016, gaiadr2-2018a} --  whose main purpose is to determine accurate parallaxes and proper motions for over one billion stars -- provided the astrophysical community with additional observational constraints to estimate the stellar age.
The Gaia RD2 catalogue \citep{gaiadr2-2018a} provides luminosities for about 100 millions stars. Thus it is worth exploring in some details if the adoption of this additional constraint, together with those adopted so far, can improve the precision  of the estimated ages and how the error in luminosity propagates in the final estimates.

Several warnings were raised by the Gaia collaboration about the adoption of the luminosities provided by the catalogue. They are due to the neglect in their derivation of both the extinction and the error in effective temperature \citep{gaiadr2-2018b}. An optimistic estimate of the average error in the provided luminosities is about 15\% \citep{gaiadr2-2018b}.

This quoted average error is much larger than those in the asteroseismic observables used in our analysis and cannot be adopted in our simulations. The computed grid of stellar models is not wide enough to explore the parameter space without suffering of severe edge effects. As a result, the estimates are severely shrunk and provide and unrealistic evaluation of the observational error impacts.
Therefore 
we decided to drastically reduce the error in $L$ -- at a level of 2.5\% -- that maybe is not realistic and not achievable but for a very small minority of stars, but is comparable with the observational errors in asteroseismic constraints. We therefore repeated the sampling and estimation procedure described in Sects.~\ref{sec:method} and \ref{sec:results} adding the luminosity to the observational constraints and linear models.

A first result is that the adoption of the luminosity to the observational constraints does not lead to higher precision on the age estimates. Even with an assumed error at 2.5\% level, the standard deviation of the estimated ages in MS and RGB phases is not improved. Indeed the standard deviation of the estimated ages changes from 0.70 Gyr without $L$ in the observation pool to 0.76 Gyr in MS and from 0.69 Gyr to 0.64 Gyr in RGB. This result is quite expected because the added constraint is strongly correlated with the asteroseismic ones, therefore providing no insight into a different evolutionary characteristic, and its error is comparable to the others. 

However the impacts of $1 \sigma$ error in each observable on the final error in the estimated age -- provided in Table~\ref{tab:contributi-gaia} -- is quite different from the ones in Table~\ref{tab:contributi}. The main changes are about the relevance of the error in effective temperature in MS, which nearly vanishes when luminosity is added into the model. Moreover, the estimated impact of $\Delta \nu$ and $\nu_{\rm max}$ in RGB vary significantly. The strong highlighted changes are not surprising because the luminosity is strongly correlated with $\Delta \nu$ and $\nu_{\rm max}$ and depends on the effective temperature through the Stefan-Boltzmann relation $L = 4 \pi \sigma_B R^2 T^4_{\rm eff}$, where $R$ is the stellar radius and $\sigma_B$ is the
Stefan-Boltzmann constant. 

Overall, the impact of the error in $L$ is very important in all the explored evolutionary phases being about $-0.3$ Gyr. 
Ultimately, the computations show how the insertion of an additional variable in the models can alter drastically the estimated impacts of all the others. This is clearly understandable for the quantities considered in this work, because either $\nu_{\rm max}$ and $L$ have an explicit dependence on $T_{\rm eff}$. The simultaneous presence of these constraints causes the models to lower the direct dependence on the effective temperature in the age estimates.  

In summary, it seems that in this case, the adoption of the luminosity from Gaia DR2 catalogue as a further constraint does not provide any improvement in the precision of the age estimates. This conclusion is strengthened by the much larger error in $L$ expected for the vast majority of Gaia stars with respect to the one considered here.

\begin{table*}[ht]
        \centering
        \caption{As in Table~\ref{tab:contributi}, but with the additional constraint on the stellar luminosity.}\label{tab:contributi-gaia}
        \begin{tabular}{lrrrrrr}
                        \hline\hline
                & ${\cal E}(T_{\rm eff})$ & ${\cal E}({\rm [Fe/H]})$ & ${\cal E}(\Delta \nu)$ & ${\cal E}(\nu_{\rm max})$ & ${\cal E}(L)$ & $\Delta \nu/\Delta \nu_{\sun}$\\ 
                \hline
                Initial MS  & 0.04 & -0.23 & -0.37 & -0.47 & -0.29 & 1.75 \\ 
                End MS & 0.05 & -0.39 & -0.04 & -0.28 & -0.38 & 0.50 \\ 
                RGB         & 0.27 & -0.06 & 0.11  & -0.48 & -0.30 &  \\ 
                \hline
        \end{tabular}
\end{table*}

\section{Conclusions}\label{sec:conclusions}

Thanks to the availability of high-quality data and the development of sound statistical approaches the problem of inferring stellar ages has been widely addressed in the recent literature. While several works explored some aspects of the propagation of the observational uncertainties to the inferred ages and on the expected biases, the direct quantification of how measurement errors in single observational constraints propagate on the final estimate is still largely unexplored. The question has indeed a relevant importance for observational purposes because it would highlight the quantities for which a refinement in the measurement would be most beneficial.

We performed a theoretical investigation on the direct impact of observational errors on the recovered age for stars in MS and RGB phases. We assumed that a mix of classical (effective temperature and metallicity [Fe/H]) and asteroseismic ($\Delta \nu$ and $\nu_{\rm max}$) constraints were available for the objects.

We focussed on a reference isochrone at metallicity [Fe/H] = 0.3 and age 7.5 Gyr. Artificial stars were sampled from the reference isochrone and subjected to random Gaussian perturbation in their observational constraints. The age of these synthetic objects were then recovered by means of a Monte Carlo Markov chains approach. 

The difference between the recovered and true ages were modelled against the offset in the observational values, either by means of linear models and by means of projection pursuit regression models. The first class of statistical models provides an easily interpretable and  generalizable result, while the second one allows a check for the robustness of the inference. In fact, due to the complex way the offset in the observational quantities influence the recovered age, the hypotheses of linear model theory are not totally fulfilled. Luckily, we find that the inference from the linear model was a good proxy for that from projection pursuit regression models. Thus the inference on the impact of $1 \sigma$ offset of the observational constraints obtained through them can be safely adopted for judging the criticality of an observational quantity.

For MS the effective temperature of the star is the most important source of variability. An offset of $+75$ K causes a change in the estimated age from $-0.4$ to $-0.6$ Gyr  in the  initial and terminal parts of the MS respectively. An observation error of $+2.5\%$ in $\nu_{\rm max}$ accounts for a change of about $-0.3$ Gyr. A 0.1 dex error in [Fe/H] results  important only at the end of the MS, causing an age error of $-0.4$ Gyr. The contribution of $\Delta \nu$ is negligible.

For the RGB phase the dominant source of uncertainty is $\nu_{\rm max}$, whose variation causes an age underestimation of about 0.6 Gyr; the offset in the effective temperature and $\Delta \nu$ account respectively for an underestimation and overestimation of 0.3 Gyr. The metallicity offset plays a minor role.

In the light of these results it is clear that no source of uncertainty is globally dominant in the whole stellar evolution to the purpose of age recovering. However, it seems that the effective temperature and the frequency of maximum oscillation power play an important role in both the explored phases. Unfortunately both these quantities are knowingly problematic either from the observational or from the theoretical point of view. In fact 
although it is not uncommon to find quoted errors in the effective temperature of the order of some tens of degrees, typically a comparison of results by different authors shows discrepancies even larger than 100 K \citep[see e.g.][]{Ramirez2005,Masana2006,Casagrande2010, Schmidt2016}.  Regarding the maximum oscillation power, its theoretical understanding is far less certain than that of $\Delta \nu$, which is inversely proportional to the sound travel time through the star and directly to the square root of the stellar mean density \citep{Ulrich1986,Kjeldsen1995}. Therefore the adoption of scaling relations to fit the data could be potentially problematic. Indeed, the validity of the scaling relation was questioned for both the adopted asteroseismic quantities in the late RGB stages \citep{Epstein2014, Gaulme2016,Viani2017}.     

We also investigate the relevance of adopting as additional observational constraint the stellar luminosity, since this quantity is provided by the Gaia DR2 catalogue.  To this purpose -- giving computational limitations -- we strongly reduced the suggested average error in the luminosity in the Gaia DR2 catalogue from 15\% to 2.5\%. Even in this very favourable configuration the adoption of the luminosity constraint does not allow better precision on the estimated ages, while changing the individual contribution of the single error sources. Overall, the impact of the adopted $1 \sigma$ luminosity error on the age estimates is about $-0.3$ Gyr for all the explored evolutionary phases. A much larger error is expected when adopting the typical precision provided by the Gaia DR2 catalogue.

While the present work sheds some light on the complex way in which individual observational uncertainties affects the stellar age estimation, it only scrapes the surface of this problem. Without direct verification it is not clear if the results of the present investigation can be generalized to other mass ranges and metallicities. While the latter hypothesis is quite probable, the former is much more questionable due to the different morphology of evolutionary tracks for heavier stellar objects, which occurs after the development of a convective core. Moreover, other evolutionary phases, such as the central helium burning stage, should be explored in similar way. Additional theoretical work is still needed before grasping a sound comprehension of the relevance of a single predictor in the estimation of stellar ages.

\begin{acknowledgements}
We thanks our anonymous referee for the stimulating suggestions and useful comments.
This work has been supported by PRA Universit\`{a} di Pisa 2018-2019 
(\emph{Le stelle come laboratori cosmici di Fisica fondamentale}, PI S.~Degl'Innocenti) and by INFN (\emph{Iniziativa specifica TAsP}).
\end{acknowledgements}

\bibliographystyle{aa}
\bibliography{biblio}

\appendix

\section{Full form of the linear models}\label{app:fullform}

The full form of linear model, with explicit coefficient dependence, for the MS phase is
\begin{eqnarray}
{\cal E}({\rm age}) \, ({\rm Gyr})  & = & \beta_0 + \beta_1 \Delta \nu * [\beta_2 {\cal E}(T_{\rm eff}) +  \beta_3 {\cal E}({\rm [Fe/H]}) + \nonumber \\ 
&+&   \beta_4 {\cal E}(\Delta \nu) + \beta_5 {\cal E}(\nu_{\rm max}) ].\label{eq:MS-model-full}
\end{eqnarray}
The $\beta$ coefficients are listed, in order, in the first column of Table~\ref{tab:MS-add}. So $\beta_0$ corresponds to the Intercept term, $\beta_1$ to the $\Delta \nu$ term and so on.  

For the RGB phase, the full model is
\begin{eqnarray}
{\cal E}({\rm age}) \, ({\rm Gyr})  & = & \beta_0 + \beta_1 {\cal E}(T_{\rm eff}) +  \beta_2 {\cal E}({\rm [Fe/H]}) + \nonumber \\ 
&+&   \beta_3 {\cal E}(\Delta \nu) + \beta_4 {\cal E}(\nu_{\rm max}),\label{eq:RGB-model-full}
\end{eqnarray}
The coefficients are listed in Table~\ref{tab:RGB-add}.

\section{PPR detailed results}\label{app:ppr}

\begin{figure*}
        \centering
        \includegraphics[height=17cm,angle=-90]{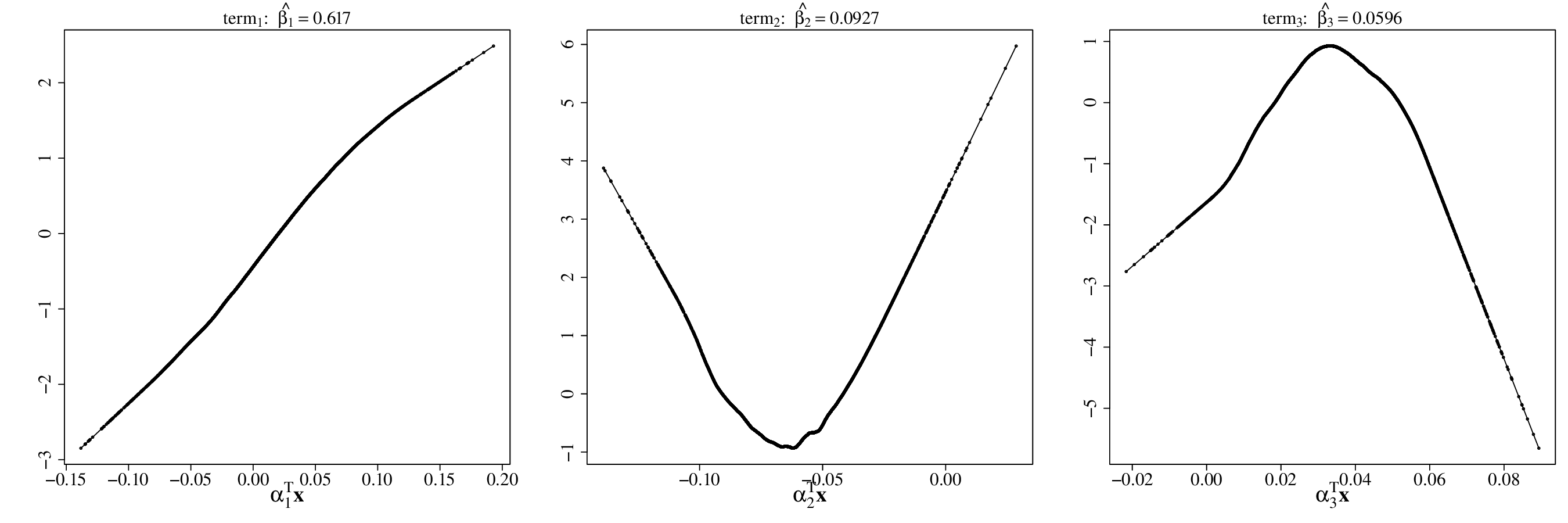}
        \includegraphics[height=10.2cm,angle=-90]{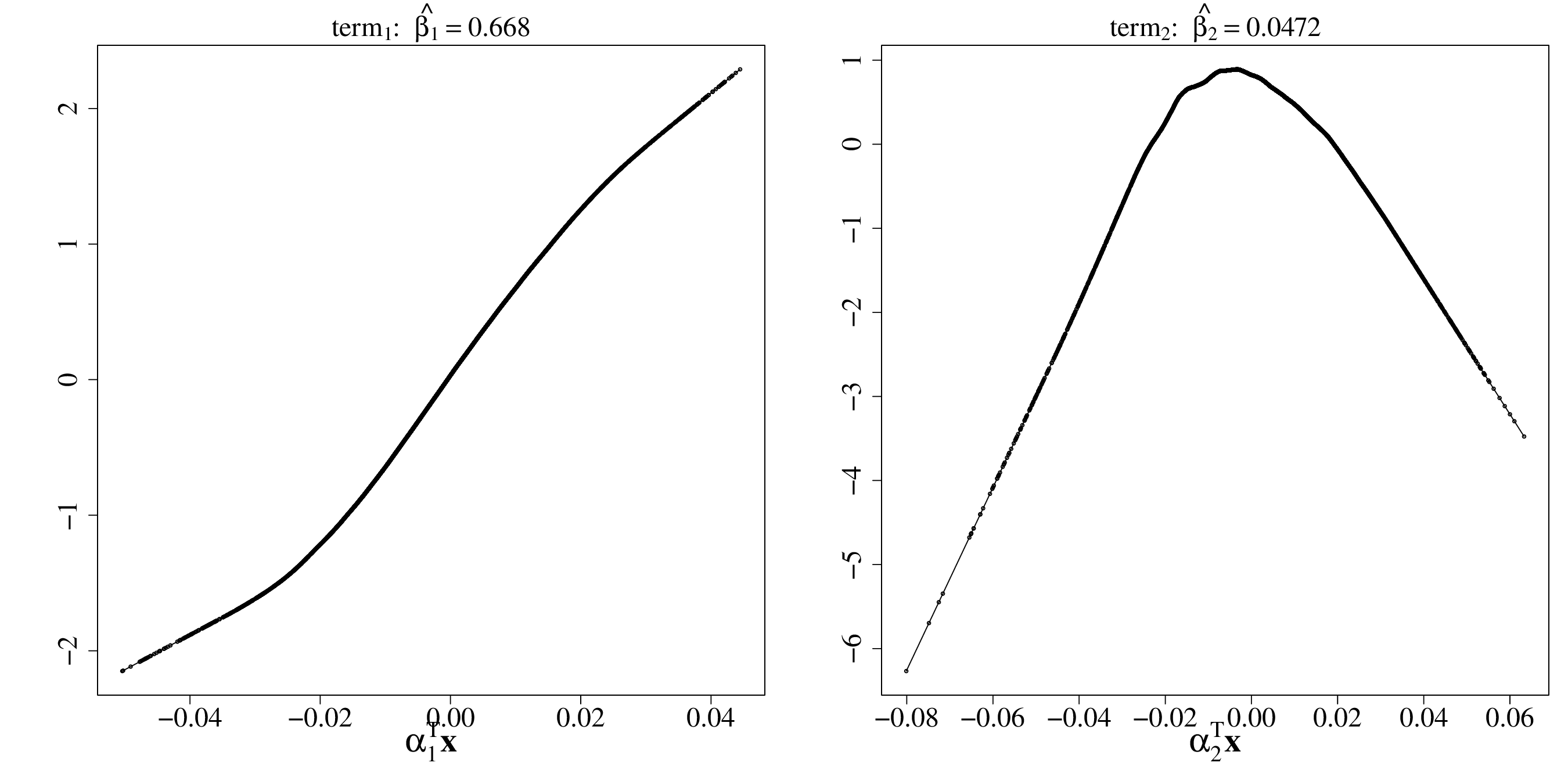}
        \caption{{\it Top} row: the three identified ridge functions for PPR models specified in Eq.~(\ref{eq:ppr1}) for the MS phase.
                The $\beta$ coefficients from the fit are also shown.  {\it Bottom} row: same as in the {\it top row} but for the two ridge function in the RGB phase.}
        \label{fig:ppr-ridge}
\end{figure*}

The coefficients of the $\alpha$ vectors in Eq.~(\ref{eq:ppr1}) are provided in Tables~\ref{tab:ppr-MS} and~\ref{tab:ppr-RGB}, while the empirical smoothing function are shown in Fig.~\ref{fig:ppr-ridge}.

\begin{table}[ht]
        \centering
        \caption{Components of $\alpha$ vectors for the PPR model of the MS phase, for the $M = 3$ fitted terms.} 
        \label{tab:ppr-MS}
        \begin{tabular}{lrrr}
                \hline\hline
                & term 1 & term 2 & term 3 \\ 
                \hline
                $\Delta \nu$ & $1.82 \times 10^{-2}$ & $-5.12 \times 10^{-2}$ & $3.01 \times 10^{-2}$ \\ 
                ${\cal E}(T_{\rm eff})$ & $-5.80 \times 10^{-4}$ & $-1.33 \times 10^{-4}$ & $-7.79 \times 10^{-5}$ \\ 
                ${\cal E}({\rm [Fe/H]})$ & $-2.15 \times 10^{-1}$ & $-1.20 \times 10^{-1}$ & $-8.93 \times 10^{-2}$ \\ 
                ${\cal E}(\Delta \nu)$ & $1.71 \times 10^{-1}$ & $9.73 \times 10^{-1}$ & $9.89 \times 10^{-1}$ \\ 
                ${\cal E}(\nu_{\rm max})$ & $-9.61 \times 10^{-1}$ & $-1.88 \times 10^{-1}$ & $-1.11 \times 10^{-1}$ \\ 
                \hline
        \end{tabular}
\end{table}

\begin{table}[ht]
        \centering
        \caption{Components of $\alpha$ vectors for the PPR model of the RGB phase, for the $M = 2$ fitted terms.} 
        \label{tab:ppr-RGB}
        \begin{tabular}{rrr}
                \hline\hline
                & term 1 & term 2 \\ 
                \hline
                ${\cal E}(T_{\rm eff})$ & $-9.12 \times 10^{-5}$ & $-1.55 \times 10^{-4}$ \\ 
                ${\cal E}({\rm [Fe/H]})$ & $6.98 \times 10^{-3}$ & $1.07 \times 10^{-1}$ \\ 
                ${\cal E}(\Delta \nu)$ & $8.02 \times 10^{-1}$ & $7.76 \times 10^{-1}$ \\ 
                ${\cal E}(\nu_{\rm max})$ & $-5.98 \times 10^{-1}$ & $-6.22 \times 10^{-1}$ \\ 
                \hline
        \end{tabular}
\end{table}

\end{document}